# Title page

# Persuasive Technology in Reducing Prolonged Sedentary Behavior at Work: A Systematic Review

Yunlong Wang [a], Lingdan Wu [b], Jan-Philipp Lange [c], Ahmed Fadhil [d], Harald Reiterer [a]

[a] HCI Group, University of Konstanz, Konstanz, Germany.

[b] Psychological Assessment and Health Psychology Group, University of Konstanz, Konstanz, Germany.

[c] Social and Health Sciences Group, University of Konstanz, Konstanz, Germany.

[d] Centro Ricerche GPI, Fondazione Bruno Kessler (FBK-irst), Trento, Italy.

yunlong.wang@uni.kn, lingdan.wu@uni.kn, jan-philipp.lange@uni.kn
fadhil@fbk.eu, harald.reiterer@uni.kn

## ABSTRACT

Prolonged sedentary behavior is prevalent among office workers and has been found to be detrimental to health. Preventing and reducing prolonged sedentary behavior require interventions, and persuasive technology is expected to make a contribution in this domain. In this paper, we use the framework of persuasive system design (PSD) principles to investigate the utilization and effectiveness of persuasive technology in intervention studies at reducing sedentary behavior at work. This systematic review reveals that reminders are the most frequently used PSD principle. The analysis on reminders shows that hourly PC reminders alone have no significant effect on reducing sedentary behavior at work, while coupling with education or other informative session seems to be promising. Details of deployed persuasive technology with behavioral theories and user experience evaluation are lacking and expected to be reported explicitly in the future intervention studies.

# 1. INTRODUCTION

## 1.1 Prevalence of Sedentary Behaviors and Associated Health Consequences

"Sitting has become the smoking of our generation (Merchant, 2013)." It has become a common sense that prolonged sedentary behavior is very unhealthy. This correlates with many preventable diseases and deaths, such as cardiovascular diseases, diabetes, obesity, high blood pressure, colon cancer, and so on (Brakenridge et al., 2016; Dustan et al., 2011; Knaeps et al., 2016; Thorp et al. 2011). Sedentary behavior (SB) refers to any waking behavior characterized by an energy expenditure less than or equal to 1.5 METs (or metabolic equivalent) with a sitting or reclining posture during tasks such as working at a desk and watching TV (SBRN, 2012). Prolonged sedentary behavior has been defined as maintaining sedentary behavior for more than 30 minutes, and this definition has been well adopted in this domain (Hadgraft et al., 2016; Henson et al., 2016).

People may think meeting physical activity guidelines (e.g., 150 minutes of moderate-intensity aerobic activity and more than 2 times muscle-strengthening activity per week) can compensate for the deleterious health consequences of prolonged SB. However, there has been evidence showing that too much sitting and too little moderate- to vigorous-intensity physical activity (MVPA) represent separate and distinct risk factors for chronic, non-communicable diseases (e.g., cardiovascular disease, diabetes, cancer) (Knaeps et al., 2016; Shrestha et al., 2016). In other words, changing the amount of physical activity cannot completely attenuate the negative health consequences of SB. Therefore, an increasing number of recent studies focused on interventions targeting the sedentary lifestyle (Chau et al., 2010; Chu et al., 2016; Healy et al., 2012; Urda et al., 2016). In this review paper, we specifically focus on the intervention studies applying persuasive technology that aimed to reduce prolonged SB at workplaces where it is very common and has caused many health problems among office workers. We will following present the related works and the reason why we are interested in the application of persuasive technology to this domain.

## 1.2 Effectiveness of Intervention Strategies

To reduce sedentary behavior and make people be more active, many intervention studies have been conducted and several review papers examining the effectiveness of the interventions are available. Nevertheless, none of them focused on the impact of technology on reducing SB.

A pioneering work by Chau (2010) addressed the effect of workplace interventions in reducing SB. It reviewed studies on the effectiveness of workplace interventions to reduce sitting behavior. However, all the selected studies focused on increasing physical activity with reducing sitting time as a side effect or a secondary goal, which may weaken the effectiveness of the interventions on reducing sedentary behavior (Gardner et al., 2016; Prince et al., 2014).

Based on 11 studies that aimed to improve health conditions at work by reducing SB, a narrative review by Healy et al. (2012) supported the use of multiple strategies (e.g., increasing the number of breaks from sitting time, changing postural, focusing on ergonomic changes to the individual workspace, altering the built design of the broader workplace, and using multiple strategies) to reduce prolonged workplace sitting, since these strategies could not only improve the participants' health conditions at work, but also typically had a beneficial or neutral impact on productivity, absenteeism and injury costs.

With a meta-analysis, Shrestha et al. (2016) aimed to provide a more precise understanding of the workplace interventions for reducing the SB at work. Eight studies were included with a total of 1125 participants who were grouped into three categories: physical workplace changes, policy changes, and information and counseling. The results showed that sit-stand desks can reduce sitting time at work, while the effects of policy changes and information and counseling are inconsistent. It was pointed out that all the reviewed studies were at high risk of bias and the quality of the evidence was low due to small sample sizes and poor research design (i.e., inadequate randomization, allocation concealment, blinding of outcome assessment, incomplete outcome data, or selective reporting).

The problem with the low quality of evaluation methods was also mentioned in the review by Gardner and colleagues. In this review, Gardner et al. (2016) focused on identifying effective behavior change strategies used in sedentary behavior reduction interventions, based on intervention functions (Michie et al., 2011) and the taxonomy of behavior change techniques (BCTs) (Michie et al., 2013). They found that self-monitoring, problem-solving, and restructuring the social or physical environment were particularly promising BCTs in reducing SB among adults.

In a recent systematic review (Chu et al., 2016), consistent evidence for intervention effectiveness was found for reducing the SB in workplace, particularly for multi-component (i.e., deploying sit-stand workstations in combination with behavioral interventions) and environmental strategies (i.e., using sit-stand workstation, portable elliptical/pedal machine, and stationary cycle ergometer and treadmill desk). Educational/behavioral strategies on their own (i.e., motivational interviewing, provide information on consequences of behavior to the individual, goal setting, action planning, prompt self-monitoring of behavior, provide instruction on how to perform the behavior, teach to use prompts/cues and facilitate social comparison), were less effective. The pooled intervention effect showed a significant workplace sitting reduction of -39.6 min/8-h workday (95% confidence interval [CI]: -51.7,-27.5), favoring the intervention group. Although this review found a significant result of effectiveness in SB reduction interventions, it did not compare the effectiveness of different behavior change techniques as shown in (Gardner et al., 2016), which is necessary for guiding intervention design.

**1.3 Persuasive Technology**

From the introduction in section 1.2, we can see the majority of interventions to reduce SB require changing environment or workplace policy, which are not available in many cases. We believe that the modern technology could provide promising tools for reducing SB in an effective and efficient way. Without a doubt, our lifestyle and daily behavior have been changed dramatically by modern technologies, such as computers and smartphones, and office workers spend most of their SB time using modern technologies and devices. Therefore, we seek ways of reducing prolonged SB from technology, which we call persuasive technology (PT).

The term, *persuasive technology*, describes technologies designed to change users' attitude and behavior (Fogg, 2002). Fogg pointed out three methods technologies can change people: as tools, as media, and as social actors. Based on this understanding, Oinas-Kukkonen and Harjumaa (2009) developed a framework called Persuasive System Design (PSD) that can be directly applied to persuasive system development. The PSD framework describes 28 persuasive technology principles in 4 categories (supporting primary task, computer-human dialogue, system credibility, and social) and explains how to transfer these principles into software functionality.

The PSD has been used explicitly and implicitly in health behavior change intervention studies (Lehto & Oinas-Kukkonen, 2015; Matthews et al., 2016). Kelders et al. (2012) provided a systematic review of the impact of the PSD on adherence to web-based interventions. Wildeboer and colleagues (2016) conducted a meta-analysis showing that web-based interventions with the principles in the PSD model have a large and significant effect size on mental health, and increasing the number of principles in different categories does not necessarily lead to better outcomes. In addition, they also found a number of combinations of principles that were more effective, e.g., tunneling and tailoring, reminders and similarity, social learning and comparison.

Unlike behavior change theories (e.g., Transtheoretical Model of Change (TTM) (Prochaska & Velicer, 1997) and the Health Belief Model (HBM) (Champion, 1984)) that explain and predict behavior, persuasive technologies describe the characteristics how information systems should deliver behavior change interventions. Another confusion would be the relationship between persuasive technology and behavior change techniques (BCTs), which is a well-known taxonomy containing 93 items (e.g., goal-setting). As rooted from behavior change theories, BCTs provide the content for behavior change interventions. Although some elements (e.g., self-monitoring and rewards) appear in both BCTs and PSD principles, they are derived from different perspectives. We believe persuasive technologies should be used with proper behavior change theories in practice.

### 1.4 Aim

To our knowledge, there has been no systematic review of the impact of persuasive technology on interventions of reducing prolonged SB at work. Our aim, therefore, is to provide exactly this review of PT in the domain of SB change at work. In the rest of this paper, we will first explain our systematic review process and the results. Then we show the analysis of the reviewed studies and the pitfalls for further studies. Finally, we give the conclusion and future work.

## 2. METHODS

### 2.1 Data Source

Articles were searched using Google Scholar[1], ACM DL[2], JMIR[3], and PubMed[4]. The terms and the term combination strategy used to search target articles are listed in Table 1. All articles were published between 1987 and November 2016. The identified articles were further manually searched for relevant publications.

### 2.2 Study Selection

Relevant papers were imported into Mendeley[5] Desktop software, and duplicates were removed. The four-phase flow diagram of PRISMA (Liberati et al., 2009) was used to illustrate the study selection process. We filtered the interventions following the criteria:

- Target Group: Only adults who have sedentary lifestyle at work.

---

[1] https://scholar.google.com

[2] http://dl.acm.org/

[3] http://www.jmir.org/

[4] https://www.ncbi.nlm.nih.gov/pubmed

[5] https://www.mendeley.com

- Target Behavior: Only interventions aiming to reduce prolonged sitting behavior at work.
- Study Design: Only studies with a parallel control group.
- Measurement: Only studies reporting sitting time output objectively measured by activity tracker or self-report.
- Language: Only articles written in English.
- Persuasive Technology: Only interventions/intervention arms integrating PSD principles as variables.

| Number | Term | Combination |
|---|---|---|
| 1 | Workplace | |
| 2 | Occupation | |
| 3 | At work | |
| 4 | Office | (1 OR 2 OR 3 OR 4) AND (5 OR 6) AND (7 OR 8 OR 9) AND 10 |
| 5 | Sedentary | |
| 6 | Sitting | |
| 7 | Adults | |
| 8 | Worker | |
| 9 | Employee | |
| 10 | Intervention/s | |

**Table 1: the terms and combination strategy for searching articles**

## 2.3 Data Coding

All the selected interventions were coded according to the PSD principles (Oinas-Kukkonen & Harjumaa, 2009), as shown in Table 2.

| PSD principle | Definition |
|---|---|
| **Primary Task Support** | |
| Reduction (1.1) | System should reduce steps users take when performing target behavior. |
| Tunneling (1.2) | System should guide users in attitude/ behavior change process by providing means for action. |
| Tailoring (1.3) | System should provide tailored info for user groups. |
| Personalization (1.4) | System should offer personalized content and services for individual users. |
| Self-monitoring (1.5) | System should provide means for users to track their performance or status. |
| Simulation (1.6) | System should provide means for observing link between cause & effect with |
|  | regard to users' behavior. |
| Rehearsal (1.7) | System should provide means for rehearsing target behavior. |
| **Dialogue Support** |  |
| Praise (2.1) | System should use praise to provide user feedback based on behaviors. |
| Rewards (2.2) | System should provide virtual rewards for users to give credit for performing target behavior. |
| Reminders (2.3) | System should remind users of their target behavior while using the system. |
| Suggestion (2.4) | System should suggest users carry out behaviors while using the system. |
| Similarity (2.5) | System should imitate its users in some specific way. |
| Liking (2.6) | System should have a look & feel that appeals to users. |
| Social role (2.7) | System should adopt a social role. |
| **System Credibility Support** |  |
| Trust-worthiness (3.1) | System should provide info that is truthful, fair & unbiased. |
| Expertise (3.2) | System should provide info showing knowledge, experience & competence. |
| Surface credibility (3.3) | System should have competent and truthful look & feel. |
| Real-world feel (3.4) | System should provide info of the organization/actual people behind it content & services. |
| Authority (3.5) | System should refer to people in the role of authority. |
| Third-party endorsements (3.6) | System should provide endorsements from external sources. |
| Verifiability (3.7) | System should provide means to verify accuracy of site content via outside sources. |
| **Social Support** |  |
| Social learning (4.1) | System should provide means to observe others performing their target behaviors. |
| Social comparison (4.2) | System should provide means for comparing performance with the performance of others. |
| Normative influence (4.3) | System should provide means for gathering people who have same goal & make them feel norms. |
| Social facilitation (4.4) | System should provide means for discerning others who are performing the behavior. |
| Cooperation (4.5) | System should provide means for cooperation. |
| Competition (4.6) | System should provide means for competing with others. |
| Recognition (4.7) | System should provide public recognition for users who perform their target behavior. |

**Table 2: PSD principles.**

## 3. RESULTS

### 3.1 Search and Selection Results

A total of 1025 articles were identified from online database searching and other records, of which 38 articles were duplicated. Two of the authors (Y.W. and L.W.) screened the records separately and then resolved the conflicts together. Afterward, 708 records were screened out since no clear information relevant to the research topic was found. Finally, eight intervention studies were selected after full-text reading (see Figure 1). One of the authors (Y.W.) coded the interventions and the other two authors (J.L. and A.F.) checked the coding list. Differences were resolved by discussion.

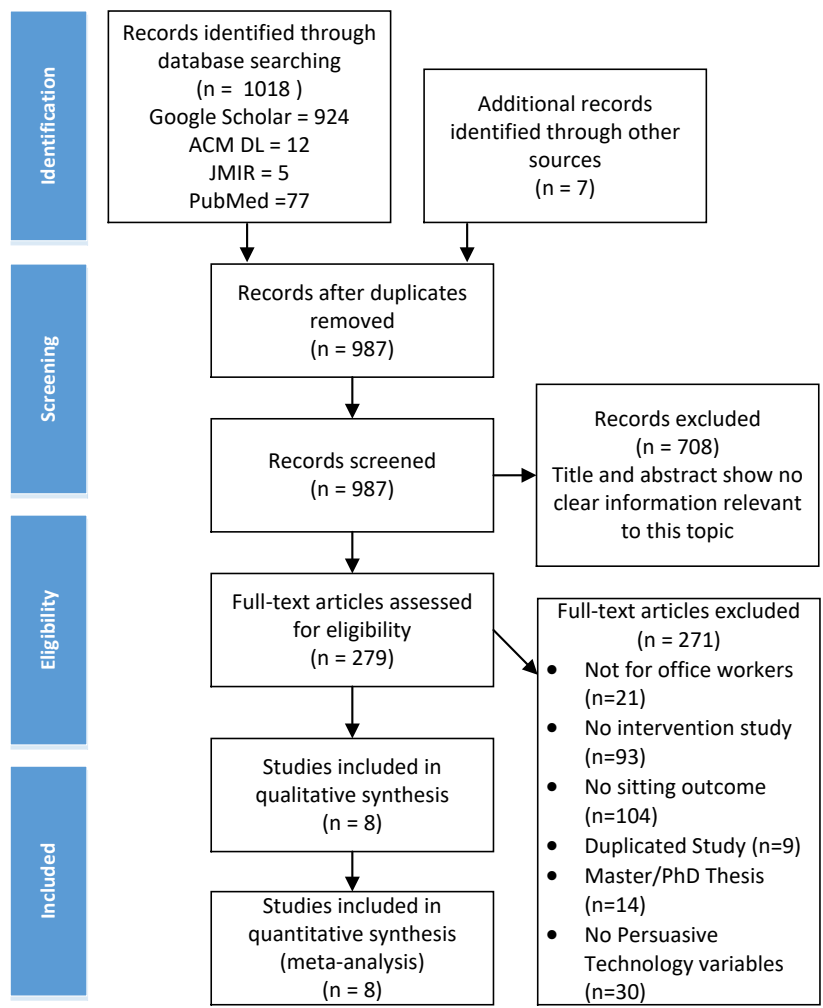

**Figure 1: The study selection workflow**

## 3.2 Intervention Studies Description

All of the selected studies were conducted in developed countries within the last five years. Six out of eight studies are Randomized Controlled Trials (RCTs), while the other two are quasi-experiments. Only participants who finished the intervention duration are included. The sample size ranged from 28 to 246 and the interventions lasted from 5 days to 12 months. SB was objectively measured using activPAL or ActiGraph activity trackers in most of the studies (n=6), and three studies used self-reporting questionnaire (IPAQ (Craig et al., 2003) and WSQ Chau et al., 2011) to estimate the physical activity and sedentary behavior. Details are listed in Table 3.

The aims of the interventions are all to reduce office workers' sedentary behavior during work and/or waking hours, but with different proximal aims, e.g., increasing standing time, short walks, or breaks. Four studies used no-treatment control conditions, while the other studies compared traditional intervention plus PT-intervention with a traditional intervention-only comparator arm. Three studies examined the effectiveness of hourly PC-based prompt to reduce sitting time at work (Donath et al., 2015; Evans et al., 2012; Wildeboer et al., 2016).

As mentioned in (Chu et al., 2016), organizational and environmental factors also impact sedentary behavior at work. Even though we focus on a specific measurement, such as total sitting time, the effect size may also have a large bias based on different intervention-control settings. Both (Donath et al., 2015) and (Evans et al., 2012) used a PC-based prompt to reduce sitting time at work. But in (Evans et al., 2012) all participants attended an education session, while in (Donath et al., 2015) participants in both groups were facilitated with height-adjustable working desks. Only three intervention studies contained a no-treatment comparator arm, which can only lead to less powerful meta-analysis.

### 3.3 Effects of Interventions

Four interventions (De Cocker et al., 2016; Evans et al., 2012; Gilson et al., 2016; Puig-Ribera et al., 2015) showed statistically significantly positive effect on reducing sedentary behavior in at least one outcome (e.g., prolonged sitting time), compared with control groups. In another study (Brakenridge et al., 2016) both the intervention group and control group showed significant improvement in SB. Among the remaining studies, one study (Donath et al., 2015) reported a notable but not significant decrease in sitting time and increase in standing time compared to the control condition, while the other one found a significant increase of weekday sedentary behavior with a small effect size in the intervention group. More details are listed in Table 4.

### 3.4 Persuasive Technology Analysis

Reminders (i.e., reminding participants of their goals to reduce the SB) was the most frequently used PSD principle among the reviewed studies, while tunneling (i.e., guiding participants to reduce SB by PC-based prompt, website, or email), self-monitoring (i.e., providing means to keep track of one's own SB in real-time), and tailoring (i.e., providing tailored information to specific participant group or intervention phase) appeared in three studies (see Figure 2). Reflecting to the four function categories, primary task support (i.e., tunneling, tailoring, personalization, and self-monitoring) and dialogue support (i.e., reminders and suggestion) were often utilized, while social support and system credibility support appeared not as frequently, as only social comparison (i.e., sharing participants' experience through social networks) and expertise (i.e., informing users about the health risks of prolonged SB) were used in these two categories.

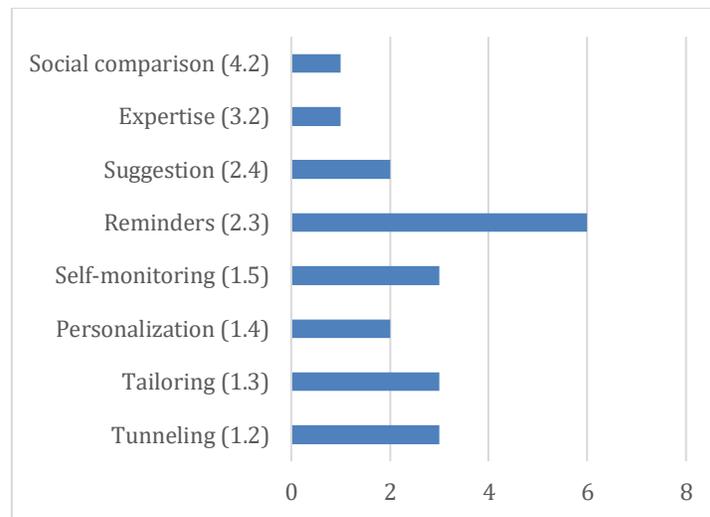

Figure 2: PSD principles (see details in Table 2) usage frequency

### 3.5 In-Depth Analysis of Reminders

As six reviewed studies used reminders to inform participants of their sedentary behavior and timely breaking moment, it is of interest to make an in-depth analysis on it. We list the details of these reminders based on the reminder properties including frequency, interface, and content in Table 5.

*3.5.1 Reminders Frequency*

Two reviewed interventions (Taylor et al., 2016; Urda et al., 2016) examined the effectiveness of hourly PC-based prompts on reducing occupational sitting along with tunneling or suggestion found no significant effect compared with the control groups. However, another intervention (Evans et al., 2012) adopted similarly frequent (every 30 min) PC reminders showed significant effect following a brief education session. Only one intervention (Gilson et al., 2016) used event-based reminders (i.e., every 30-60 min continuous sitting), which also conducted a participatory workshop before the intervention and resulted with a significant and larger effect on reducing SB than controlled group.

*3.5.2 Prompts Interface*

In most of the interventions (Donath et al., 2015; Evans et al., 2012; Gilson et al., 2016; Tylor et al., 2016; Urda et al., 2016), reminders were delivered via PC prompts with text messages, while only one via the smartphone (Brakenridge et al., 2016). The effect of the interface cannot be evaluated based on the results of the reviewed studies because there are no enough controlled conditions to compare the different effect of this variable.

*3.5.3 Reminders Contents*

The contents of the reminders are not only the plain text to remind the time but also more complex sentences which reflect the other principles like suggestions (Donath et al., 2015; Tylor et al., 2016) and changing appearance like color (Gilson et al., 2016). However, the contents seem to have no effect on the interventions because different contents lead to same results when comparing (Tylor et al., 2016) and (Urda et al., 2016). It is important to note that all of the contents in the reviewed studies are fixed, from which the users can get less information after several times appearance of the reminders.

## 4 PITFALLS

### 4.1 PSD Principles Overlapping

When coding the intervention elements into PSD principles, we found that it is difficult to differ some similar meaning pairs, e.g., tailoring (1.3) and personalization (1.4), tunneling (1.2) and suggestion (2.4). The tailoring principle was defined as "Information provided by the system will be more persuasive if it is tailored to the potential needs, interests, personality, usage context, or other factors relevant to a user group," while the personalization was "a system that offers personalized content or services has a greater capability for persuasion." So it is clear that the personalization should be a case of tailoring. Similarly, the tunneling principle and the suggestion principle are also overlapping. These overlaps can lead to confused coding. In this paper, we regard the property allowing users to modify the interface as personalization and code multiple suggestions which can be followed step by step into tunneling.

| Author, year, reference No. | Country | Study design | Sample size (finished) | Study duration | Measurement method | Intervention-control Description | Intervention Aim |
|---|---|---|---|---|---|---|---|
| Evans et al. (2012) | UK | RCT | 28 (I:14 C: 14) | 5 work days for baseline, 5 work days for intervention | objectively measured using the activPAL activity monitor | C: a brief education session on the importance of reducing long sitting periods at work<br>I: same education along with prompting software on PC reminding to stand up every 30 minutes. | To reduce long uninterrupted sedentary periods and total sedentary time at work. |
| Donath et al. (2015) | Switzerland | RCT | 31 (I: 15 C: 16) | 5 work days for baseline, 12 weeks for intervention | objectively measured by ActiGraph wGT3X-BT | C: height-adjustable working desk<br>I: height-adjustable working desk and three daily screen-based prompts | To increases daily office standing time in healthy middle-aged office workers. |
| Puig-Ribera et al. (2015) | Spain | Quasi-experimental | 190 (I:88 C:102) | 5 work days for baseline, 19 weeks for intervention, 2 weeks for follow-up | objectively measured by pedometers, self-reported using IPAQ | C: daily logging of their steps and sitting time<br>I: daily logging and a web-based program with automatic email delivering in a three-phase intervention | To decrease occupational sitting time through increased incidental movement and short walks. |
| Brakenridge et al. (2016) | Australia | RCT | 153 (I: 66 C: 87) | 12 months | objectively measured using ActivPAL monitors | C: only organizational support<br>I: organizational support plus LUMOback activity tracker | To reduce sitting in office workers. |
| Taylor et al. (2016) | USA | RCT | 106 (I: 59 C: 47) | 6 months | self-reported using IPAQ | C: no intervention<br>I: PC-based prompt | To improve physical and behavioral health outcomes. |
| De Cocker et al. (2016) | Belgium | RCT | 93 (I1: 35 I2: 35 C: 23) | 3 months | self-reported using WSQ, objectively measured using activPAL activity monitor | C: no intervention<br>I1: automatic Web-based, generic information and suggestions<br>I2: automatic Web-based, computer-tailored intervention | To reduce and interrupt sitting at work. |
| Gilson et al. (2016) | Australia | Quasi-experimental | 57 (I: 24 C: 33) | 5 months | objectively measured using GENEActive activity monitor | C: participatory workshop only<br>I: participatory workshop and a chair sensor/software package (Sitting Pad) with real-time prompts | To reduce occupational sedentary exposure and increase physical activity. |
| Urda et al. (2016) | Australia | RCT | 44 (I: 22 C:22) | 2 weeks | objectively measured using activPAL 3 | C: no intervention<br>I: maintained behaviors during control week, but received hourly alerts on their computer during work hours in the intervention week | To reduce sitting time, increase sit-to-stand transitions, and improve perceived wellness in women with sedentary jobs. |

**Table 3: Intervention study characteristics**

| Author, year, reference No. | PSD principle | Outcomes | Intervention Group Results | Controlled Group Results |
|---|---|---|---|---|
| Evans et al. (2012) | 2.3 | Total sitting time (minutes/day [%]) Number of sitting events (events/ day [events/hour]) Number of prolonged sitting events (events/day [events/hour]) Duration of prolonged sitting events (hours/day [%]) | Total sitting time reduction is not significate, but the prolonged sitting time reduction is significate (-48(84) min/day or -12.2% (19.3%) with 95% CI, p<0.05). | No significant difference |
| Donath et al. (2015) | 1.2  2.3 | Standing time (hours per week) Sitting time (hours per week) | Standing time 7.2 (4.8) to 9.7 (6.6) hours/week with p=0.09, sitting time 29.4 (6.5) to 27.8 (10.7) hours/week with p=0.63. | No significant difference |
| Puig-Ribera et al. (2015) | 1.2  1.3 1.5  2.4 4.2 | Step (counts/day) Sitting time (minutes/day) | Daily occupational sitting reduced significantly -32.2(9) min/day p=0.007, and step counts increased +924(245) P<0.001. | No significant difference |
| Brakenridge et al. (2016) | 1.5  2.3 | Work hours and Overall hours: Sitting (min/10 h workday) Prolonged sitting (min/10 h workday) Time between sitting bouts (min) Standing (min/10 h workday) Stepping (min/10 h workday) Step count (number of steps/10 h workday) | Only in work hours, significant improvement in sitting (−35.5 (25.3) min, p = 0.006), prolonged sitting (−45.7 (38.3) min, p = 0.019), standing (+27.4 (19.7) min, p = 0.007), and stepping (+9.1 (8.9) min, p = 0.045). | During work hours: significant improvement in sitting (−40.5 (20.4) min, p < 0.001), p = 0.006), prolonged sitting (−41.3 (26.5)min, p = 0.002), standing (+39.2 (18.3) min, p < 0.001), and Time between sitting bouts (+1.7 (1.4) min, p = 0.019) |
| Taylor et al. (2016) | 1.2  2.3 | Weekday Sedentary time (min/weekday) Weekend Sedentary time (min/weekend) | Weekday sedentary behavior has no significant change (P = .20), weekend sedentary behavior decreased significantly. | Weekday sedentary behavior increased significantly, weekend sedentary behavior has no significant change. |
| De Cocker et al. (2016) | 1.3  1.4 1.5  2.4 3.2 | Self-report Total sitting time (minutes/day) Self-report domain-specific sitting (minutes/day) objectively measured total sitting time awake (hours/day) objectively measured sitting at work (% work hours) objectively measured standing at work ((% work hours) objectively measured breaks at work (No./work hours) | Significant decrease in self-reported total sitting (507 (104) to 425 (110) min/day, P<0.001) and sitting at work (338 (107) to 259 (88) min/day, P<0.001), and significant increase in objectively measured breaks at work (3.8(1.5) to 4.3(1.6) hours/day, P=0.09). | No significant difference |
| Gilson et al. (2016) | 1.3  1.4 2.3 | Sedentary behavior in % work time Light physical activity in % work time Moderate physical activity in % work time Total time sitting (min/day) Longest bout sitting (min/day) | Significant decrease in Sedentary (-8%, p = 0.012), increase in light physical activity (+8%, p=0.018), and decrease in longest bout sitting (-15 min). | Significant decrease in sedentary (-2%), increase in light physical activity (+1%) and increase the longest bout sitting (+17 min) |
| Urda et al. (2016) | 2.3 | during an 8.5-hour workday: Time sitting (hours/workday) sit-to-stand transitions | No significant difference in average sitting time and sit-to-stand transitions from baseline compared with intervention. | No significant difference in average sitting time and sit-to-stand transitions from baseline compared with intervention. |

**Table 4: Interventions effect**

| Study | Frequency | Interface | Content |
|---|---|---|---|
| Evans et al. (2012) | Every 30 min. | PC prompt with text (enforced showing 1 min each time). | "It's a break time." |
| Donath et al. (2015) | Three fixed times per day. | PC prompt with text (could be closed manually) | "Prolonged sitting is harmful!; Change your working position!; Lift up your working desk" |
| Brakenridge et al. (2016) | Not mentioned. | Smartphone prompt | Not mentioned. |
| Taylor et al. (2016) | Hourly. | PC prompt | Tips to encourage users to get up and walk hallways, stairs, or outdoors. |
| Gilson et al. (2016) | Every 30-60 min continuous sitting. | PC prompt with a color indicator (from green to amber, and then to red). | Changing color. |
| Urda et al. (2016) | Hourly | PC prompt with audible alert. | "Get up and move." |

**Table 5: Reminder properties in the reviewed interventions.**

## 4.2 Behavioral Theories

We regard sedentary behavior reduction as a health behavior change problem. However, among the selected intervention studies, only one (De Cocker et al., 2016) reported that it was based on behavior change theories (i.e., self-determination theory (Deci & Ryan, 2002), the theory of planned behavior (Ajzen, 1991), and self-regulation theory (Maes & Karoly, 2005)). Since evidence has shown that behavior change interventions based on theories of health behavior are more effective than the non-theory-based ones (Davis et al., 2015; Glanz & Bishop, 2010), more theory-integrated sedentary behavior reduction studies are expected. According to the related reviews (Davis et al., 2015; Glanz & Bishop, 2010), the most frequently used theories of behavior change include the Transtheoretical Model of Change (TTM) (Prochaska & Velicer, 1997), the Health Belief Model (HBM) (Champion, 1984), the Social Cognitive Theory (SCT) (Bandura, 1977), and the Theory of Planned Behavior (TPB) (Ajzen, 199), to name a few. We would encourage more research applying these theories or novel theories into this domain.

## 4.3 User Experience

No study measured user experience of the deployed persuasive technology after the interventions. The reason may be that most of the SB reduction studies were conducted by non-HCI researchers. However, it is essential to take the user experience into consideration when evaluating the persuasive technology.

The user experience of short-term and long-term usage of the persuasive technology may reveal mediator effects between user acceptance and intervention effectiveness.

## 5 IMPLICATIONS

Based on our review results, we provide the following implications:

(1)    Well-designed intervention studies (e.g., RCTs) on reducing prolonged sedentary behavior at work with explicit involvement of persuasive technology are still lacking. Therefore, we encourage interdisciplinary cooperation in this field.

(2)    When applying PSD principles in intervention studies, the underlying behavior change theories are supposed to be reported explicitly. As social support in reviewed studies are rarely applied, it should be further explored in future studies.

(3)    Researchers of persuasive technology or PSD should pay attention to the user experience of behavior change support systems that apply persuasive technologies.

(4)    Based on the reviewed studies, only using PC-prompt reminders with tunneling or suggestions show no significance on reducing sedentary time, while combined with brief education session the reminders can significantly improve sedentary behavior. This can be a good option for corresponding public intervention designers and providers.

## 6 STRENGTHS AND LIMITATIONS

This systematic review is the first to examine the effectiveness of persuasive technology on reducing prolonged sedentary behavior. We also show that there is a lack of theory-based interventions and user experience considerations in the selected studies. However, given the small number of studies and inconsistent study design, we did not conduct a meta-analysis to analyze the correlation between Persuasive System Design principles and intervention outcomes.

## 7 CONCLUSION AND FUTURE WORK

Through the systematic review of the intervention studies to reduce sedentary behavior at work, we illustrated how the persuasive technology was used. We revealed that reminders are the most frequently used PSD principle, while the principles of system credibility support and social support were seldom deployed. An analysis based on the frequency, the interface and the contents of the reminders gave more insights on how was this most popular persuasive technology utilized. We also showed the pitfalls of the

PSD principles and the reviewed interventions studies including the behavioral theories basis and user experience evaluation.

More intervention studies are expected to explicitly report the details following the PSD principles to make more powerful systematic review and meta-analysis. The research on the user experience of the persuasive technology delivered to reduce prolonged sedentary behavior at work should draw more attention from the HCI community. More theory-based behavior change interventions utilizing persuasive technology are required to enable comprehensive meta-analysis. More longitudinal studies are also required to evaluate the long-term effects of SB reduction interventions.

The future work would be more focused on the properties of the variety of persuasive technologies to design better interventions on reducing sedentary behavior.

## ACKNOWLEDGMENTS

This work was partially supported by SMARTACT. Y. Wang acknowledges the financial support from the China Scholarship Council (CSC).

## CONFLICT OF INTEREST STATEMENT

None.

A meta-analysis. International journal of medical informatics, 96, 71-85. https://doi.org/10.1016/j.ijmedinf.2016.04.005.